# Multichannel Design of Non uniform Constellations for Broadcast/Multicast Services

Belkacem Mouhouche, Mohammed Al-Imari, Daniel Ansorregui
*Samsung Electronics Research & Development UK, Staines-Upon-Thames, TW18 4QE, UK*
{b.mouhouche, m.al-imari, d.ansorregui}@samsung.com

***Abstract*—Recent studies have shown the potential performance gain of Non Uniform Constellations (NUC) compared to the conventional uniform constellations. NUC can be a promising candidate in 5G systems to increase the data throughput. In the literature, NUC is designed for a specific SNR value and propagation channel. However, in broadcast/multicast services, the received signal by different users will see independent and different channels. Hence, in this paper, we focus on the potential gain of NUC when jointly optimized for more than one propagation channel. In order to assess the gain, we propose an iterative algorithm to jointly optimize the NUC for different channel conditions. The resulting constellations are then compared to uniform constellation and single channel NUC. The simulation results show that the newly designed constellations outperform the classical single channel NUC across different channel conditions when the average performance (across different channels) is considered.**

***Index Terms*— 5G Physical Layer, Constellations, Non-uniform QAM, broadcast/multicast.**

## I. Introduction

5th Generation (5G) cellular systems are being developed within different research and standardizations initiatives. 5G systems will support a large spectrum of use cases going from low delay low data rate to very high data rate. The delivery of a common content to a large number of receivers is a very important use case of 5G systems [1]. The common content can be accessible by all receivers in the network (broadcast) or by a subset of receivers (multicast). In order to achieve high data rates, all the possible enhancements are being considered. The Bit Interleaved Coded Modulation (BICM) chain was adopted in binary communication systems as an efficient scheme to approach the Shannon limit with affordable complexity. The BICM chain includes mainly a Forward Error Correction (FEC) encoder, a bit interleaver, and a bit to constellation symbol mapper. Up to the fourth generation (4G), the bit to symbol mapper is a simple uniform Quadrature Amplitude Mapper (QAM) mapper. Uniform QAM constellations are easy to map and de-map. However, there is no information theory basis for this choice and these constellations can be shown to be far from the Shannon limit.

In the research community, Non-Uniform Constellation (NUC) has received attention as a tool to improve the performance of uniform constellations with minor decoder complexity increase. In the NUC the constellation points are no further required to be in a uniform or rectangular shape. In [2], the author provided a first insight into the potential benefits of NUC and its impact on the BICM channel capacity. In [3], the authors present a performance analysis for their optimized NUC. The effectiveness of NUC was briefly tested in [4] followed by an extensive analysis of a large range of modulation and coding schemes in [7]. The previous analysis in [4, 7] focused on the additive white Gaussian noise (AWGN) channel scenario.

The work carried out up to now on NUC has always concentrated on the single channel BICM capacity optimization. Most of the time, the BICM capacity is optimized to perform well for the AWGN channel. The design can be performed as well using the Rayleigh channel as a basis by using the Probability Density Function (p.d.f.) of the Rayleigh distribution. However, the broadcast/multicast signal is received by receivers with different channel conditions. Thus, it is important that the selected NUC has a good performance across different channels and different signal-to-interference ratio (SNR) waterfalls. To the best of our knowledge, the analysis of the multichannel performance has not been addressed yet.

In this paper, we propose a novel algorithm to design multichannel NUC and analyze its performance compared to the single channel NUC and uniform constellations. It is shown that, on average, this method outperforms the single channel single SNR design method used up to now. Simulation results show that, compared to the uniform constellation, a potential improvement of up to 1.1 dB for 256QAM without additional complexity is expected.

The rest of the paper is organized as follows: The BICM capacity and NUC optimization are presented in Sec. II and Sec. III, respectively. In Sec. IV, we propose the multichannel NUC design algorithm. In Sec. V, we evaluate the link-level performance of the proposed multichannel NUC under different channel assumptions. Finally, concluding remarks are drawn in Sec. VI.

## II. BICM Capacity

The channel capacity of a communication link is the maximum mutual information between the channel input and output. The Gaussian distribution achieves the maximum mutual information, and the mutual information for the Gaussian distribution, also commonly referred to as the Shannon capacity, is given by [5]:

$$C = \log_2(1 + S), \tag{1}$$

where $S$ is the received SNR given by the ratio between the average received power $P$ and the noise power $N_0$. Although it is a capacity achieving distribution, the Gaussian distribution is not possible to be implemented in reality. In communication systems, more practical finite symbol alphabet channel inputs are implemented, such as QAM. The BICM capacity that characterizes the capacity of such system is given by [6]:

$$C = M - \sum_{m=1}^{M} E_{b,y}\left[log_2 \frac{\sum_{x_l \in X} p\left(\frac{y}{x_l}\right)}{\sum_{x_l \in X_b^m} p\left(\frac{y}{x_l}\right)}\right], \quad (2)$$

where $M$ is the number of the constellations bits, $y$ represents the received signal, $p(y/x_l)$ is the transition probability density function (p.d.f.) of transmitting $x_l$ and receiving $y$. $\boldsymbol{X}_b^m$ is the subset of the alphabet $\boldsymbol{X}$ (all the possible values $x_l$ $(l = 1, \ldots, N)$ constellations) for which bit label $m$ is equal to $b$.

The constellation consists of $N$ constellation points $((M{=}\log_2(N))$. The power of the alphabet of the transmitted symbols is normalized as follows:

$$P_{total} = \frac{1}{N}\sum_{l=1}^{N}|x_l|^2 = 1. \quad (3)$$

For an AWGN channel, the only parameters that affect the transition probabilities $p(y/x_l)$ are: the SNR and the constellation positions. For the Rayleigh channel, the BICM capacity is calculated across the probability distribution function of the channel.

The most straightforward way to design the alphabet $\boldsymbol{X}$ is to create uniform constellations by mapping the points $x_l$ to uniform positions. This is conventionally assumed in state of the art communication systems. However, two intuitive questions arise from this choice: how far is the BICM capacity of uniform QAM from the channel capacity (Shannon limit). Second, if the gap is significant, which optimal constellation achieves the smallest gap?

The BICM capacity of uniform QAM can be calculated from (2) and the resulting shortfall of the uniform QAM BICM capacity from the Shannon limit is shown in Fig. 1. The first question can be answered from Fig. 1 where a significant gap between BICM capacities with uniform QAM and the Shannon limit is observed. This gap increases with the constellation order. For example, a difference of 0.4 bps is observed for 256-QAM at 20 dB SNR. This gap represents the shaping gap due to the two constraints imposed by assuming uniform constellations, namely the rectangular shape and the equally spaced levels (uniform QAM).

## III. Non Uniform Constellations

In order to answer the second question and find optimal constellations that reduce the shaping gap, non-uniform constellations can be obtained by optimizing the alphabet in order to maximize the BICM capacity (2) subject to the power constraint (3).

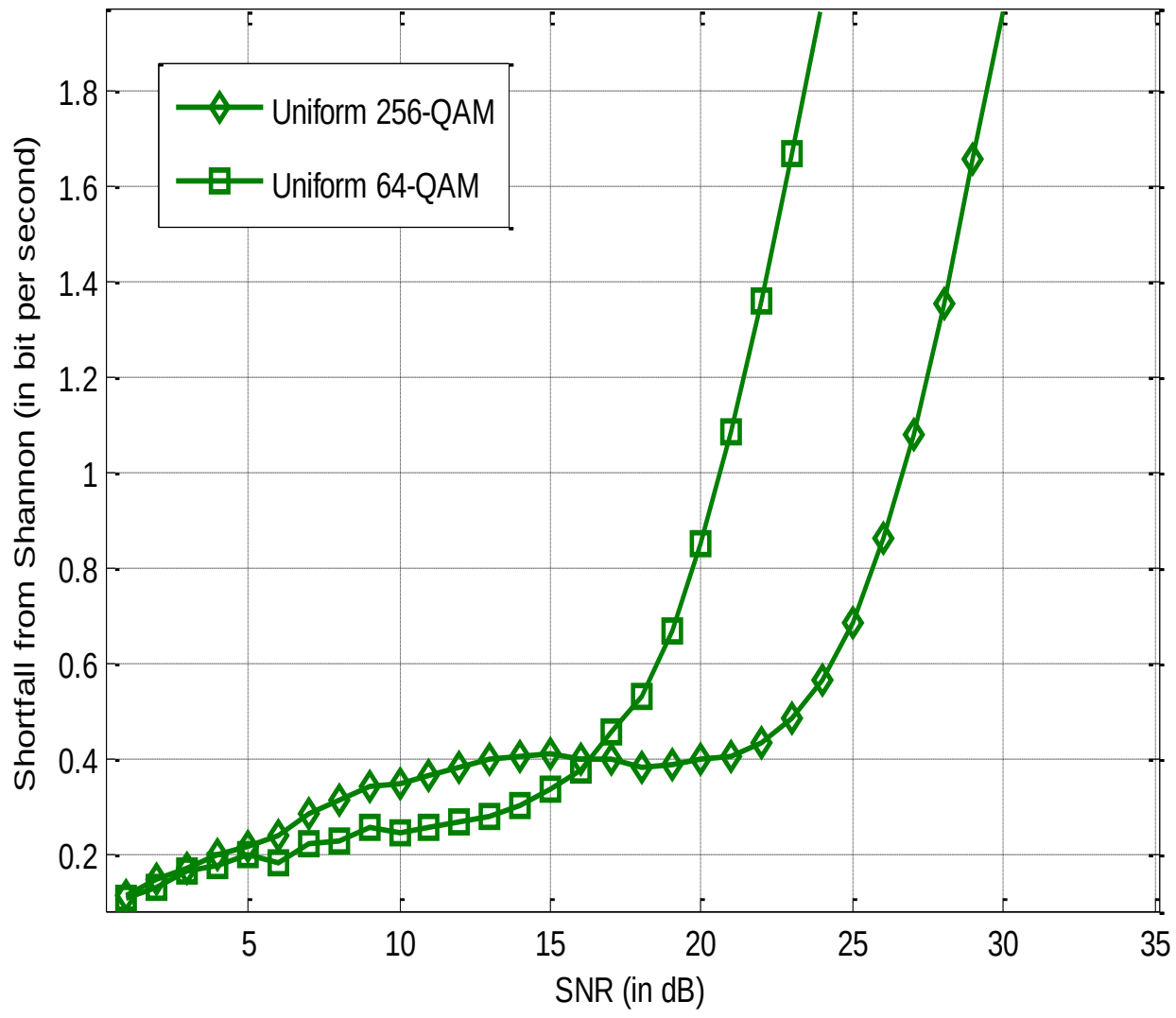

Fig. 1. Shortfall from Shannon of uniform constellations.

The optimization can be carried out using extensive search for low-order modulations. High order modulations may require the use of more sophisticated optimization methods. In this paper, we focus on how to apply the optimization for a real life BICM system. The exact capacity optimization algorithm is beyond the scope of this paper and will be discussed in a forthcoming paper. Note that since the SNR affects the transition probabilities $p(y/x_l)$ the optimal constellation will be different for each SNR.

### A. One dimensional Non-Uniform QAM (1D NUQAM)

The first type of optimization is carried out by relaxing the uniformity constraint while keeping the rectangular structure of the constellation. The advantage of this approach is twofold: on one hand the optimization is simplified because of the limited number of parameters to be optimized (Degrees of Freedom (DOF)); on the other hand the receiver can de-map the real and imaginary parts of a QAM constellation symbol independently, thus reducing the complexity of the de-mapper at the receiver end. In the sequel we will refer to this approach as the one Dimensional Non Uniform QAM (1D NUQAM). The number of DOF of 1D NUQAM is:

$$DOF_{1D\ NUQAM} = \frac{\sqrt{N}}{2} - 1. \quad (4)$$

The term $\sqrt{N}$ in (4) is due to the rectangular structure of the constellation: the optimal levels on the real and imaginary axes are equal. The factor $\frac{1}{2}$ is because the optimization is carried out on the positive levels only (the negative levels are identical). The -1 term is due to the power normalization constraint: if all the levels are fixed except one, then the remaining level can be deduced using the power constraint (4). Another way to tackle this issue is to fix the first level to 1, optimize all the remaining levels and normalize the power at the end [2]. An example of a 256 1D NUQAM constellation optimized at 11 dB is given in Fig. 2.

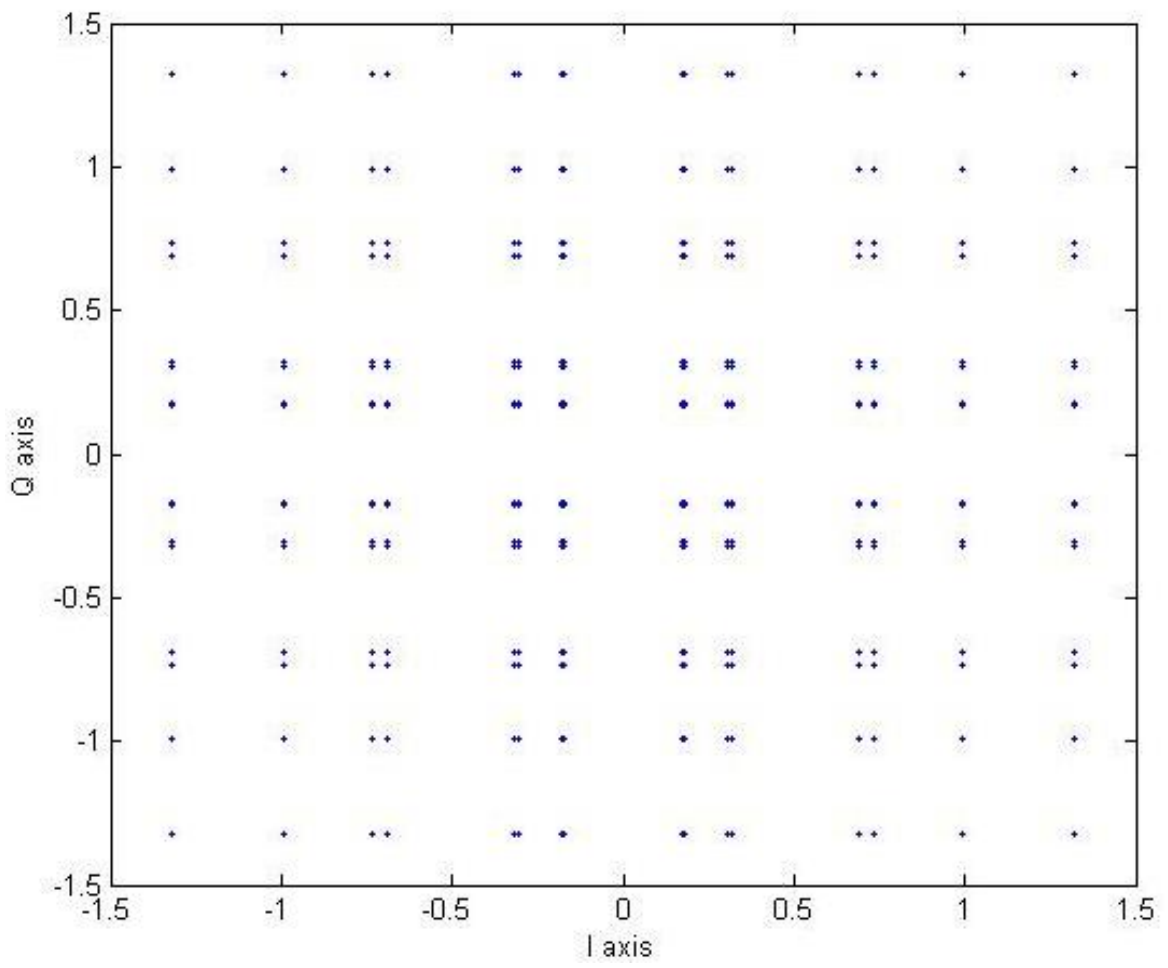

Fig. 2. 1D 256-NUQAM constellation optimized at 11dB.

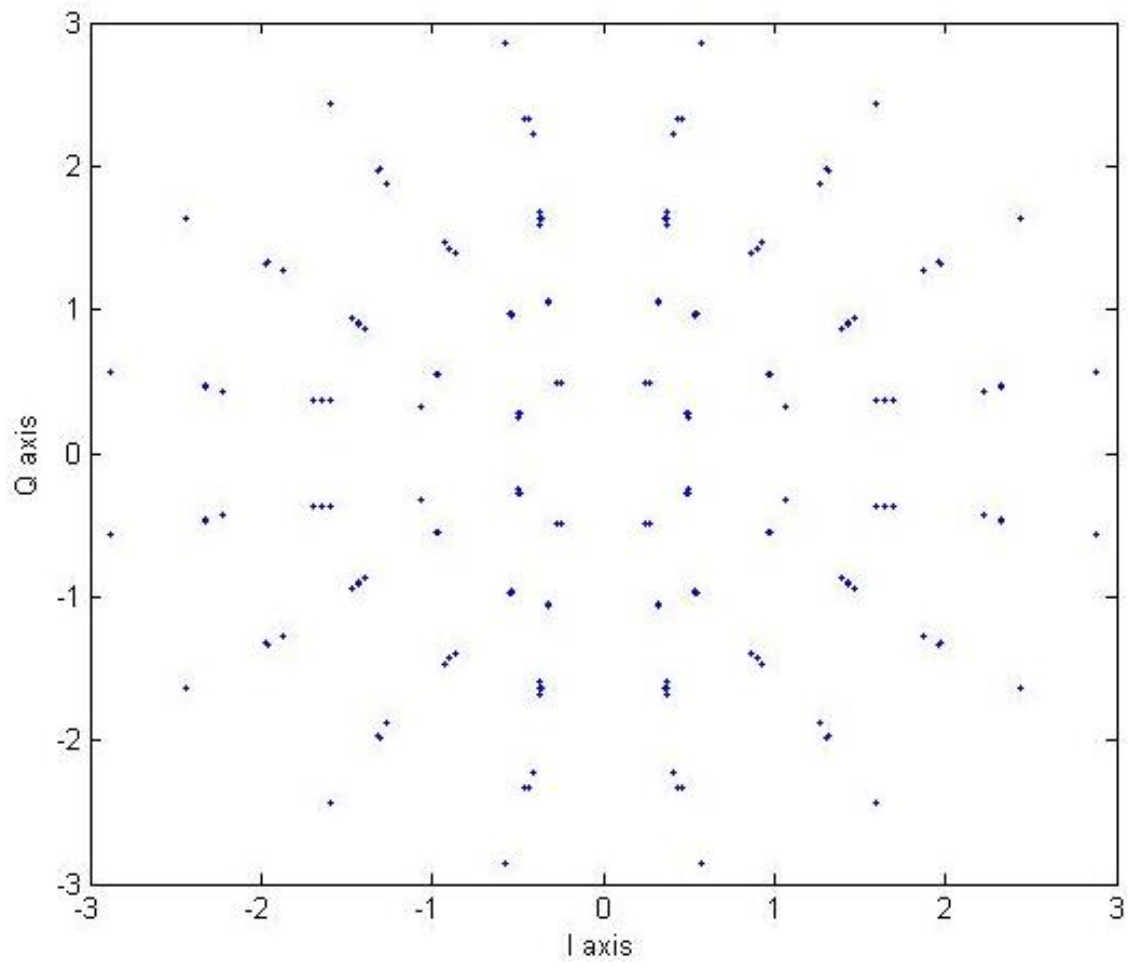

Fig. 3. 2D 256-NUQAM constellation optimized at 11dB.

### *B. Two dimensional Non-Uniform QAM (2D NUQAM)*

In order to further reduce the gap between the Shannon limit and the BICM capacity we choose to optimize the constellations by relaxing the rectangular shape constraint. In this case, the constellation values can take any shape inside one quadrant. The other three quadrants are derived from the first quadrant by symmetry. We will refer to this approach as the two Dimensional Non Uniform QAM (2D NUQAM) or simply NUC. Note that, in this case, the receiver needs to use a 2 dimensional de-mapper requiring a higher complexity than the one dimensional de-mapper. The number of DOF of 2D NUQAM is given by:

$$DOF_{2D\ NUQAM} = 2\left(\frac{N}{4}\right) - 1 \qquad (5)$$

where the factor $1/4$ is due to the fact that the four quadrants are symmetric, the factor 2 is due to the fact that the real and imaginary parts of each constellation point are optimized separately and the -1 term is due to the power normalization (if all the parameters are defined except one, then this parameter can be found by satisfying the power constraint). An example of 256 2D NUQAM optimized at 11 dB is given in Fig. 3. We see from Fig. 3 that the unconstrained optimized constellations have a circular shape although not with a uniform radius everywhere.

The Shortfall from Shannon of the 2D NUQAM and 1D NUQAM optimized constellations for 64-QAM and 256-QAM is shown in Fig. 4. We can see that, unlike Fig. 1 for uniform QAM, the shortfall for non-uniform QAM is significantly reduced. For example, the shortfall to Shannon in the SNR region 14 to 18 dB is reduced from 0.4 dB to around 0.15 dB for 2D 256-NUQAM. The same observation is valid for 1D NUQAM albeit with a slightly lower gain than 2D NUQAM.

It is clear from Fig. 4 that the BICM capacity provided by the 2D-NUQAM is higher than the BICM capacity provided by 1D-NUQAM. For this reason, we will focus on 2D-NUQAM when assessing the potential performance improvements using the multichannel optimized NUC.

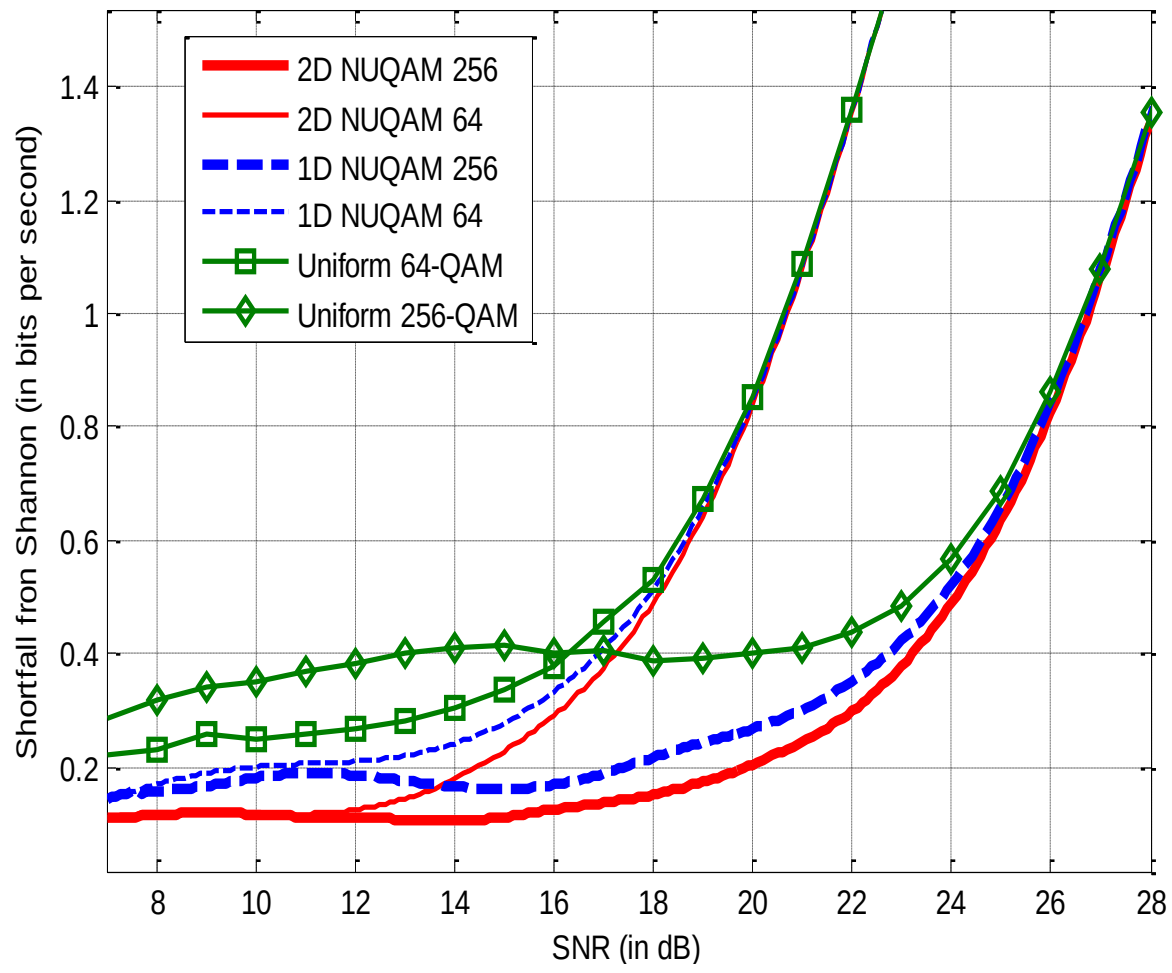

Fig. 4. Shortfall from Shannon of optimized 1D and 2D NUQAM.

## IV. NUC MULTICHANNEL DESIGN

The NUC design in the previous section is optimized for a specific SNR value. The SNR is needed in order to optimize the BICM function. However, communication systems support different coding rates for the same QAM size. Each coding rate results in a different waterfall SNR. The straightforward solution is to decouple different code rates by designing a different NUQAM for each coding rate. The design can start at the waterfall of the uniform constellation, the resulting NUC is then applied to the system and a new waterfall is obtained. This process is repeated until we obtain the final waterfall SNR and NUQAM. We have already addressed this problem in [4]. The second issue is that, in broadcast/multicast services, the transmitted signal is received by different receivers that see different channels. For example one receiver can receive a signal that propagated through an AWGN channel whereas another receiver receives a signal that propagated through a Rayleigh channel. The difficulty in this case, as opposed to the different coding rate case, is that it is not possible to decouple different channels design. This implies that the constellation needs to be the same. In order to solve this issue, we propose an algorithm that takes into account more than one channel and waterfall SNR in the design.

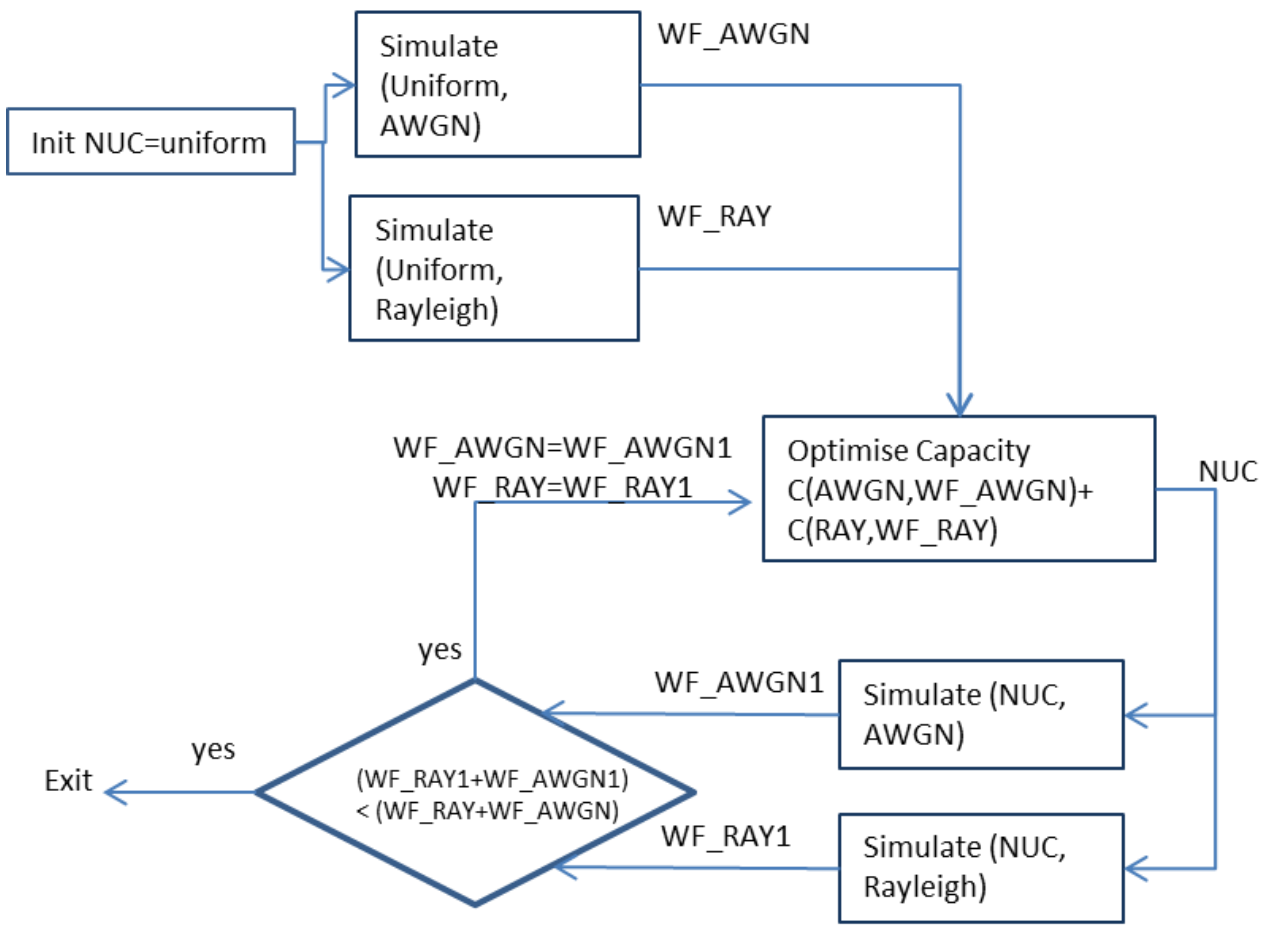

Fig. 5. Multichannel design of NUQAM.

A block diagram illustrating the algorithm's steps is shown in Fig. 5. The algorithm is initialized by setting the constellation to a uniform constellation. A simulation is then carried out to find the waterfall SNR of AWGN (WF_AWGN) and Rayleigh (WF_RAY), or any other set of wireless channels. These waterfall SNR values are then used to optimize the sum BICM capacity (sum of Rayleigh BICM capacity at the WF_RAY SNR and the AWGN BICM capacity at the WF_AWGN SNR). This step results in a new constellation NUC. This constellation NUC is then used to find the new waterfall SNRs. This process is repeated until the waterfall SNRs average stops improving. This approach will insure that the designed NUC will perform well under different wireless channel conditions.

## V. Simulation Results

In this section, the performance of the proposed NUC design algorithm is evaluated through Monte-Carlo simulations. We consider a system with 8 MHz bandwidth, FFT size of 8192 and guard interval equal to 1/8 [6].

### *Results for 64QAM*

We start by studying the performance of 64QAM with coding rate CR=3/5. For this, we design three types of NUC: i) NUC AWGN to maximize the BICM capacity for the AWGN channel at the AWGN waterfall SNR, ii) NUC Rayleigh to maximize the BICM capacity for the Rayleigh channel at the Rayleigh waterfall SNR, iii) Multichannel NUC for optimized for AWGN and Rayleigh based on the proposed algorithm in the previous section. Fig. 6 shows the bit error rate (BER) results versus SNR (in dB) for the simulated cases.

It can be observed from the figure that the performance of uniform constellation is around 0.45dB worse than the performance of AWGN NUC.

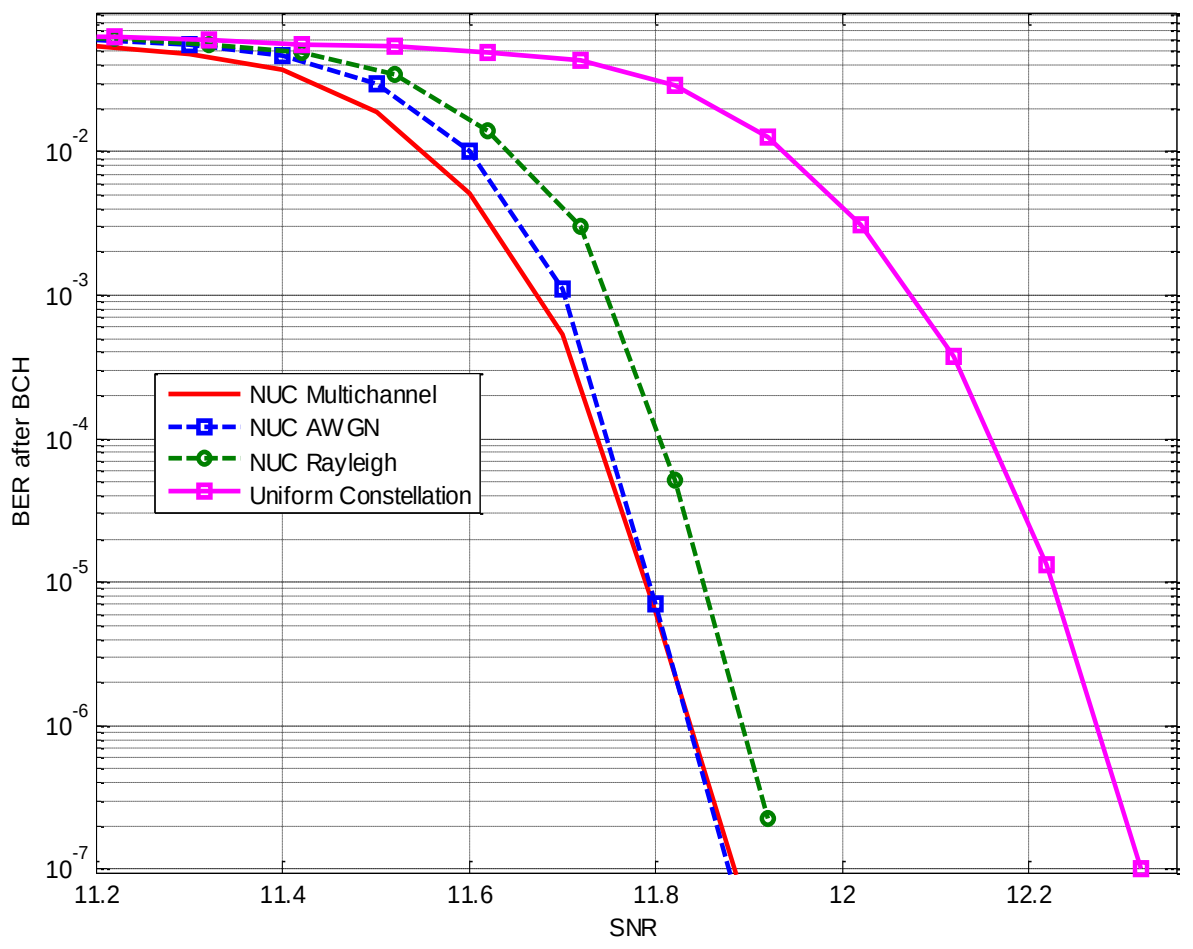

Fig. 6. Performance of different 64QAM NUCs in AWGN CR=3/5.

The performance of Rayleigh NUC, even though the channel in this case is AWGN, is still 0.35dB better than the uniform constellations. The multichannel NUC performance is almost identical to the AWGN performance.

We then conduct the same simulation for the Rayleigh channel, where the results are shown in Fig.7. The same conclusions of AWGN apply in this setting. The performance of Rayleigh optimized NUC is 0.25dB better than uniform constellations. AWGN NUC is around 0.15dB better than uniform constellations. The multichannel NUC is almost equivalent to the Rayleigh channel NUC.

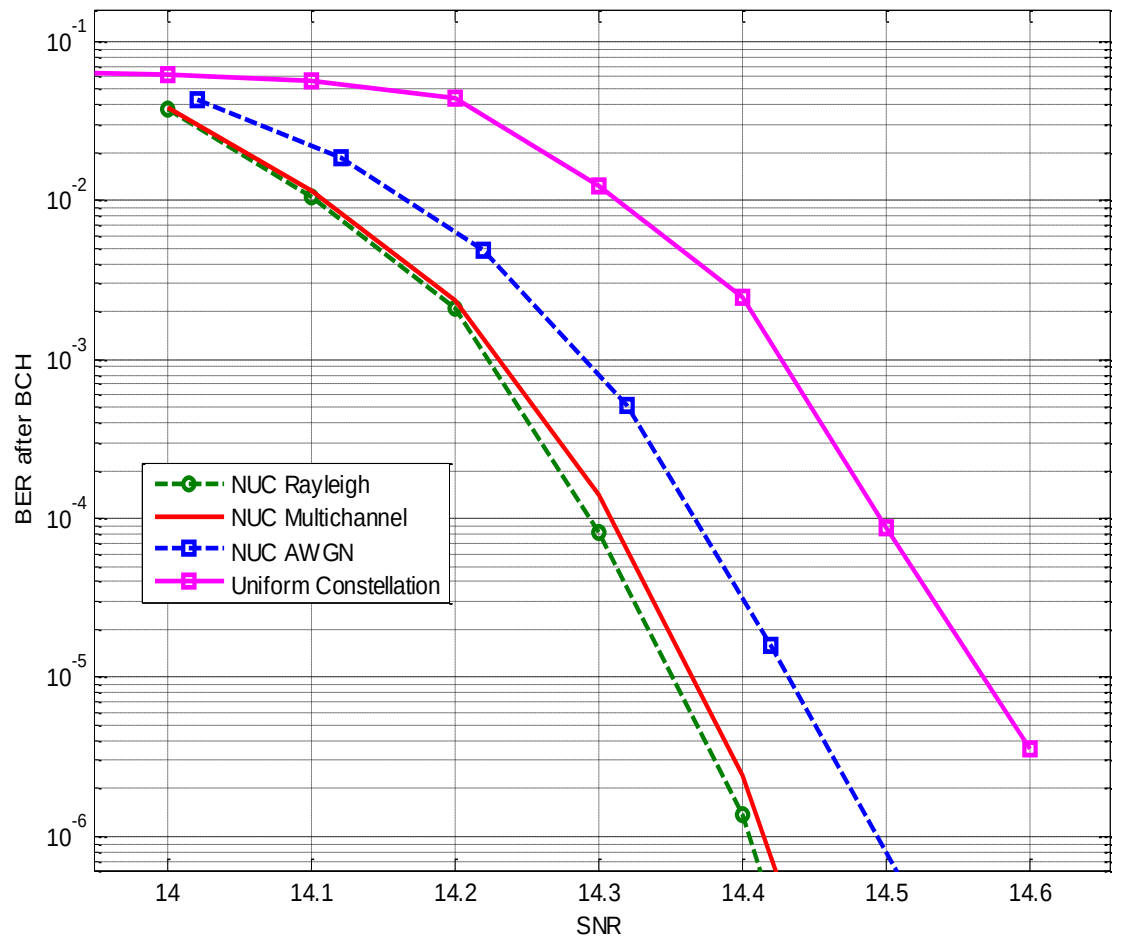

Fig. 7. Performance of different 64QAM NUCs in Rayleigh CR=3/5.

From Figs. 6 and 7 we can see that, on average, the performance of multichannel NUC is the best because it provides results that are close to the optimal case in both cases (AWGN and Rayleigh). On the other hand, the Rayleigh and AWGN designs are good when the channel of simulations matches the channel of design. However, for opposite channels (i.e. Rayleigh NUC under AWGN and vice versa) the performance is 0.1dB worse. If we suppose that different receivers receive the same signal with different channels then the use of multichannel design is better on average.

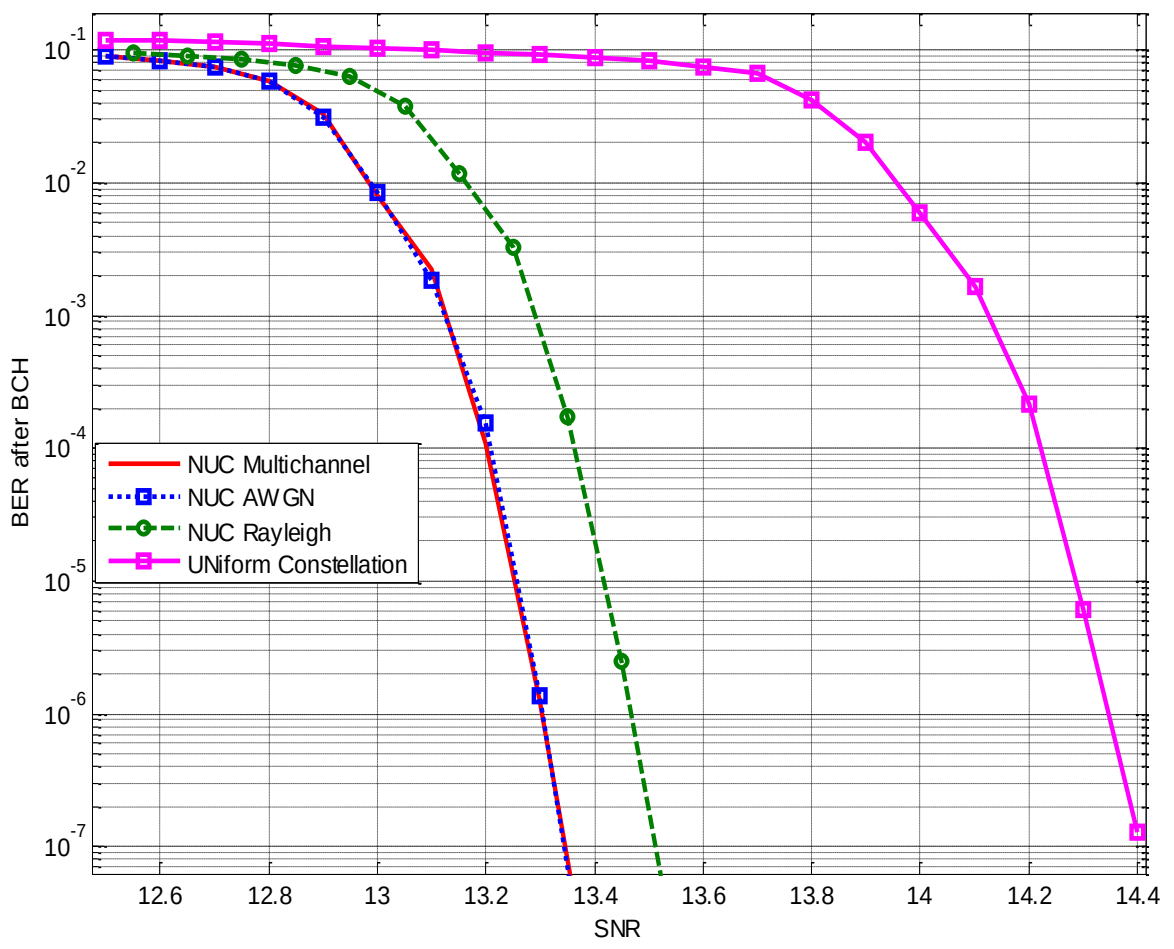

Fig. 8. Performance of different 256QAM NUCs in AWGN CR=1/2.

### *Results for 256QAM*

We then repeat the same experiment in the previous subsection with a higher order constellation of 256QAM. We also consider a different coding rate CR=1/2. The results for the AWGN channel are shown in Fig. 8.

It can be observed from the figure that the performance of AWGN NUC is 1.05dB better than the uniform constellation. The performance of Rayleigh NUC is 0.15dB worse than the AWGN one. On the other hand, the multichannel NUC provides performance identical to the AWGN performance. We next run the same experiment in Rayleigh channel. The results are shown in Fig. 9. We see from Fig. 9 that the performance of Rayleigh NUC is 0.7dB better than the uniform constellation. The performance of AWGN NUC is around 0.1dB worse than Rayleigh but still 0.6 dB better than the uniform constellation. The multichannel design proves to be very effective since the performance identical to the Rayleigh channel NUC.

Although the simulations have been conducted for AWGN and Rayleigh, the algorithm is not limited to these channels, and any arbitrary channel models can be adopted in to design the constellation.

## VI. Conclusions

In this paper, we proposed a new strategy to design Non Uniform Constellations for broadcast/multicast services in future wireless systems. The proposed method is to jointly optimize the BICM capacity and Signal to Noise Ratio for more than one channel. The proposed method results in a constellation that is close to optimal for different channels separately while giving the optimal performance on average. In the case of 256QAM it was shown that the proposed method can give a gain of around 0.15dB with respect to the NUC designed for only one channel. The same conclusions apply for 64QAM with a gain of 0.1dB. The proposed constellations do not incur any increase in demodulation complexity and require a marginal increase in the computations during the design phase. In the future we will explore the performance of this kind of design when applied to channels that have a big gap in the Waterfall SNR (AWGN, Rayleigh and TU6 channels for example).

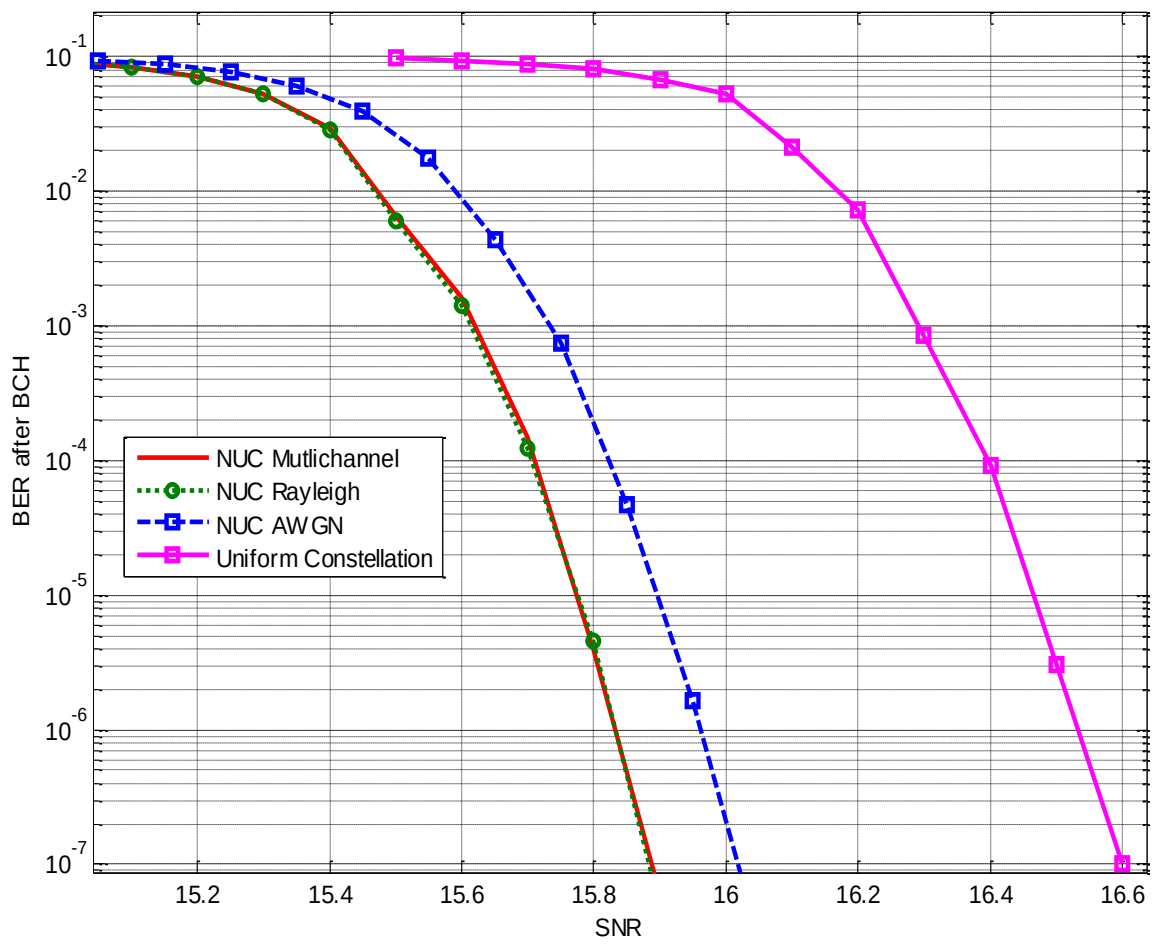

Fig. 9. Performance of different 256QAM NUCs in Rayleigh CR=1/2.

## Acknowledgment

This work has been partly performed in the framework of the Horizon 2020 project FANTASTIC-5G (ICT-671660) receiving funds from the European Union. The authors would like to acknowledge the contributions of their colleagues in the project, although the views expressed in this contribution are those of the authors and do not necessarily represent the project.